\documentclass[aps,twocolumn,preprintnumbers,showpacs,showkeys,nofootinbib%
]{revtex4}
\usepackage{epsfig}
\usepackage{amssymb,amsmath,amsfonts,amsthm,graphicx,psfrag}
\graphicspath{{./Figures/}}                 
\usepackage{array}
\usepackage{picture}
\usepackage{subfig}
\usepackage{graphicx}
\usepackage{textpos}

\newcommand{\noi}{\noindent}
\newcommand{\beq}{\begin{equation}}
\newcommand{\eeq}{\end{equation}}
\newcommand{\bea}{\begin{eqnarray}}
\newcommand{\eea}{\end{eqnarray}}

\newcommand{\Tab}[1]{Table~\ref{#1}}

\newcommand{\aabf}{{\mathbf{A}}}

\newcommand{\re}{\operatorname{\mathfrak{Re}}}
\newcommand{\tr}{\operatorname{Tr}}

\newcommand{\ds}{\displaystyle}

\newcommand{\aasymt}{{\cal A}}

\begin{document}

\title{The $\big\langle A^2 \big\rangle$ Asymmetry and Gluon Propagators\\
 in Lattice $SU(3)$ Gluodynamics at $T\simeq T_c$.}

\author{V.~G.~Bornyakov}
\affiliation{Institute for High Energy Physics NRC ``Kurchatov Institute'', 142281 Protvino, Russia \\
Institute of Theoretical and Experimental Physics NRC 
``Kurchatov Institute'', 
117259 Moscow, Russia}

\author{V.~A.~Goy}
\affiliation{{Institut Denis Poisson CNRS/UMR 7013, Universit\'e de Tours, 37200 France} \\
Pacific Quantum Center, Far Eastern Federal University, Sukhanova 8,  690950 Vladivostok, Russia}

\author{V.~K.~Mitrjushkin}
\affiliation{Joint Institute for Nuclear Research, 141980 Dubna, Russia }

\author{R.~N.~Rogalyov}
\affiliation{Institute for High Energy Physics NRC ``Kurchatov Institute'', 142281 Protvino, Russia \\
}


\begin{abstract}
We study numerically the chromoelectric-chromomagnetic
asymmetry of the dimension two $A^2$ gluon condensate
as well as the infrared behavior of the gluon propagators 
at $T\simeq T_c$ in the Landau-gauge $SU(3)$ lattice gauge theory.
We find that a very significant correlation of
the real part of the Polyakov loop with the asymmetry as well as 
with the longitudinal propagator makes it possible to determine
the critical behavior of these quantities. 
We obtain the screening masses in different Polyakov-loop 
sectors and discuss the dependence of 
chromoelectric and chromomagnetic interactions 
of static color charges and currents on the choice 
of the Polyakov-loop sector in the deconfinement phase.
\end{abstract}



\maketitle

\section{Introduction}
\label{sec:introduction}

It is widely hoped that the behavior of the Green's functions 
of gauge fields encodes the confinement mechanism  
\cite{Alkofer:2006fu,Fischer:2006ub,Huber:2018ned}.
Thus their dependence on the volume, temperature and momentum
at temperatures close to the confinement-deconfinement transition
attracts particular interest.

In the one-gluon exchange approximation, the 
Fourier transform of the gluon propagator 
measures interaction potential between static color charges.
We also should mention the relation of the low-momentum
longitudinal and transverse propagators to the 
chromoelectric and chromomagnetic screening masses
and, therefore, to the properties 
of strongly interacting quark-gluon matter.
Motivation for the studies of the asymmetry and gluon propagators
is also discussed in 
\cite{Maas:2011se,Aouane:2011fv,Chernodub:2008kf,Vercauteren:2010rk} 
and references therein.

Recently, significant correlations between 
the chromoelectric-chromomagnetic asymmetry and the Polyakov loop
as well as between the zero-momentum longitudinal propagator 
and the Polyakov loop were found in SU(2) gluodynamics  
\cite{Bornyakov:2016geh,Bornyakov:2018mmf}.
This made it possible to describe critical behavior
of the asymmetry and the propagator and to reliably
evaluate finite-volume effects.

Our attention here is concentrated on the behavior of these quantities
in the Landau-gauge $SU(3)$ lattice gauge theory.
It is well known that the first-order phase transition
occurs in this model and the Polyakov loop 
jumps from zero to a  nonzero value \cite{Yaffe:1982qf,Svetitsky:1982gs}
which is associated with the spontaneous breaking of the 
$Z_3$ center symmetry.

Though the behavior of the asymmetry and the gluon propagators 
at $T\sim T_c$ have received much attention in the literature,
the situation with their temperature and volume dependence 
in a close vicinity of $T_c$ is far from being clear.
The behavior of the gauge-field vector potentials under the 
$Z_3$ symmetry transformation is also poorly understood.

We suggested a new approach to the studies of the 
longitudinal propagator at zero momentum
$D_L(0)$, which makes it possible to clarify
its critical behavior in the infinite-volume limit of the
$SU(2)$ gluodynamics and evaluate the respective critical 
exponent to 6-digit precision \cite{Bornyakov:2018mmf}.
It was shown that the critical exponent $\gamma$ is unrelated
to the critical behavior of $D_L(0)$.

Our approach is based on correlations between 
the Polyakov loop ${\cal P}$ and $D_L(0)$ 
and between ${\cal P}$ and the asymmetry ${\cal A}$.
In the studies of these correlations
we employ well-established properties of the Polyakov loop.

The paper is organized as follows. In the next section we introduce the definition and describe the details of our numerical simulations. 
The correlation between the asymmetry ${\cal A}$ and 
the Polyakov loop ${\cal P}$ forms the subject of Section~3.
Our analysis begins with the observation that,
in a finite volume, the values of ${\cal P}$ are distributed
in a finite range making it possible to collect 
a sufficient number of configurations
generated at different temperatures, but
giving the same value of the Polyakov loop.
This allows us to study the dependence of conditional 
distributions of ${\cal A}$ on the temperature
and conclude that such distributions are governed by the value of the
real part of the Polyakov loop rather than by the temperature itself.
This finding and the knowledge of the critical behavior 
of the Polyakov loop enables one to determine the 
critical behavior of the asymmetry.
The propagators are studied in a similar way in Section~4.
We obtain the critical behavior of only the longitudinal propagator
because it correlates with the real part of the Polyakov loop
much more significantly than the transverse propagator.
Therewith, we evaluate both chromoelectric and chromomagnetic
screening masses in all Polyakov-loop sectors
and obtain their dependence on the temperature.
It turns out that, in the deconfinement phase,
the chromoelectric screening mass depends crucially on 
the choice of the Polyakov-loop sector. We discuss
consequences of this in the context of the center-cluster
scenario of the deconfinement transition \cite{Gattringer:2010ug}.
In Conclusions we summarize our findings.

\section{Definitions and simulation details}

We study SU(3) lattice gauge theory with 
the standard Wilson action in the Landau gauge. 
Definitions of the chromo-electric-magnetic 
asymmetry and the propagators
can be found e.g. in 
\cite{Chernodub:2008kf,Bornyakov:2016geh,Bornyakov:2011jm,Aouane:2011fv}.

We use the standard definition of 
gauge vector potential $\aabf_{x\mu}$ lattice \cite{Mandula:1987rh} :
\beq\label{eq:a_field}
\aabf_{x\mu} = \frac{1}{2i}\Bigl( U_{x\mu}-U_{x\mu}^{\dagger}
\Big)_{\rm traceless} \equiv A_{x,\mu}^a T^a ~,
\eeq

Transformation of the link variables $U_{x\mu}\in SU(3)$ 
under gauge transformations $g_x \in SU(3)$ has the form 
$$ U_{x\mu}
\stackrel{g}{\mapsto} U_{x\mu}^{g} = g_x^{\dagger} U_{x\mu} g_{x+\mu}\;.
$$

The lattice Landau gauge condition is given by
\beq
(\partial \aabf)_{x} = \sum_{\mu=1}^4 \left( \aabf_{x\mu}
- \aabf_{x-\hat{\mu};\mu} \right)  = 0 \,.
\eeq
It represents a stationarity condition for the gauge-fixing functional
\beq\label{eq:gaugefunctional}
F_U(g) = \frac{1}{4V}\sum_{x\mu}~\frac{1}{3}~\re\tr~U^{g}_{x\mu} \;,
\eeq
with respect to gauge transformations $g_x~$.

\vspace*{2mm}

The bare gluon propagator $D_{\mu\nu}^{ab}(p)$ is defined as
\beq
D_{\mu\nu}^{ab}(p) = \frac{a^2}{g_0^2}
\Big\langle \widetilde{A}_{\mu}^a(k) \widetilde{A}_{\nu}^b(-k) \Big\rangle~,
\label{eq:gluonpropagator}
\eeq
where $\widetilde{A}(k)$ is the Fourier transform of the
gauge potentials (\ref{eq:a_field}). 
The physical momenta $p$ are given by $p_i=\big(2/a\big) \sin{(\pi
k_i/N_s)}, ~~p_{4}=(2/a) \sin{(\pi k_4/N_t)}, ~~k_i \in (-N_s/2,N_s/2], k_4 \in
(-N_t/2,N_t/2]$. We consider only soft modes $p_4=0$.  

The gluon propagator on an asymmetric lattice 
involves two tensor structures \cite{Kapusta}:
\beq
D^{ab}_{\mu\nu}(p)\;=\;\delta_{ab} \Big( P^T_{\mu\nu}(p) D_{T}(p) +
P^L_{\mu\nu}(p) D_{L}(p)\Big)\,,
\eeq
where the longitudinal $P^{L}_{\mu\nu}(p)$ and the transverse $P^{T}_{\mu\nu}(p)$ projectors are defined at $p_4=0$ as follows:
\bea
&& P^L_{44}(p) = 1~,\quad P^L_{\mu i}(p) = P^L_{i \mu}(p) = 0 \,;\\ \nonumber
&& P^T_{ij}(p)=\left(\delta_{ij} - \frac{p_i p_j}{\vec{p}^2} \right), 
\quad P^T_{\mu 4}(p)=P^T_{4\mu}(p)=0. \nonumber
\eea
Therefore, the longitudinal $D_{L}(p)$ and
the transverse $D_T(p)$ form factors (also referred to as the longitudinal and 
transverse propagators) are given by
\beq
D_L(p) = \frac{1}{8}\sum_{a=1}^{8} D_{44}^{aa}(p);
\quad  D_T(p) = \frac{1}{16} \sum_{a=1}^{8}\sum_{i=1}^{3}D_{ii}^{aa}(p).
\eeq

\noi At $\vec{p} = 0$ the zero-momentum propagators $D_{T}(0)$ and $D_L(0)$ have the form
\beq
D_T(0) = {1\over 24} \sum_{a=1}^8 \sum_{i=1}^{3} D_{ii}^{aa}(0)\; ;
\quad D_L(0) = {1\over 8} \sum_{a=1}^8 D_{00}^{aa}(0)\; .
\eeq
The longitudinal propagator $D_T(p)$ is associated with the electric
sector and the transverse propagator $D_L(p)$ is associated with the magnetic sector.

Our calculations are performed on asymmetric lattices $N_t\times N_s^3$,
where $N_t$ is the number of sites in the temporal direction
(in our study, $N_t=8$ and $N_s = 24$).
The temperature $T$ is given by $~T=1/aN_t~$ where $a$
is the lattice spacing.We use the parameter
\beq
\tau = {T-T_c \over T_c}
\eeq 
useful at temperatures close to $T_c$.
We rely on the scale fixing procedure proposed in \cite{Necco:2001xg}
and use the value of the Sommer parameter $r_0=0.5$~fm as in \cite{Bornyakov:2011jm}. 
Making use of $\beta_c=6.06$ and 
$\ds {T_c\over \sqrt{\sigma}}=0.63$
 \cite{Boyd:1996bx} gives $T_c=294$~MeV and 
$\sqrt{\sigma}=0.47$~GeV.

In \Tab{tab:statistics} we provide information on lattice spacings,
temperatures and other parameters used in this work.

\begin{table}[tbh]
\begin{center}
\vspace*{0.2cm}
\begin{tabular}{|c|c|c|c|c|} \hline
           &          &               &                &          \\[-2mm]
 ~~~~$\beta$~~~~ & ~~$a$~fm~ & $a^{-1}$,~GeV & $p_{min}$,~MeV & ~~~~$\tau$~~~~ \\[-2mm]
           &          &               &                &          \\
   \hline\hline
 6.000 & 0.093 & 2.118 & 554.5 & -0.096 \\
 6.044 & 0.086 & 2.283 & 597.7 & -0.026 \\
 6.075 & 0.082 & 2.402 & 628.8 &  0.025 \\
 6.122 & 0.076 & 2.588 & 677.5 &  0.104 \\
\hline\hline
\end{tabular}
\end{center}
\caption{Parameters associated with lattices under study
}
\label{tab:statistics}
\end{table}

In order to consider all three Polyakov-loop sectors in detail,
we generate ensembles of $200$ independent Monte Carlo
gauge-field configurations 
for each of the sectors:
\bea
(I)   \qquad -\;{\pi\over 3} < &\arg {\cal P}& < {\pi\over 3}   \\ \nonumber
(II)  \qquad\quad    {\pi\over 3} < &\arg {\cal P}& < \pi   \\ \nonumber
(III) \qquad -\;{\pi}        < &\arg {\cal P}& < -\;{\pi\over 3} \;.   \nonumber
\eea
Consecutive configurations (considered as 
independent) were separated by $200\div 400$ 
sweeps, each sweep consisting of one local heatbath update followed
by $N_s/2$ microcanonical updates. 

Following Refs.~\cite{Bornyakov:2011jm,Aouane:2011fv} 
we use the gauge-fixing algorithm 
that combines $Z(3)$ flips for space directions with the simulated annealing 
(SA) algorithm followed by overrelaxation. 

Here we do not consider details of the approach to the continuum limit 
and renormalization considering that the lattices with $N_t=8$ 
(corresponding to spacing $a\simeq 0.08$~fm at $T \sim T_c$) 
are sufficiently fine.

In terms of lattice variables, the asymmetry has the form
\beq
\aasymt={6 a^2 N_t^2\over \beta }
\sum_{b=1}^8\left( \Big\langle A_{x,4}^b A_{x,4}^b \Big\rangle - {1\over 3}\sum_{i=1}^3
 \Big\langle A_{x,i}^b A_{x,i}^b \Big\rangle \right),
\eeq
It can also be expressed in terms of the gluon propagators:
\bea\label{eq:average_bare_asymmetry}
&& \aasymt  = {16 N_t \over \beta a^2 N_s^3}
\Big[ 3 (D_L(0) - D_T(0))  \\ \nonumber
&+& \sum_{p\neq 0}\left(
{3|\vec p|^2 \,-\, p_4^2\over p^2}  D_L(p) - 2  D_T(p)\right)\Big] \nonumber
\eea
where $D_L (D_T)$ is the longitudinal (transversal) gluon propagators.
Thus the asymmetry $\aasymt$, which is nothing but
the vacuum expectation value of the respective composite operator,
is multiplicatively renormalizable and its renormalization factor
coincides with that of the propagator\footnote{Assuming
that both $D_L(p)$ and $D_T(p)$ are renormalized by the same factor.}.
\vspace{2mm}

\section{$A^2$ asymmetry near $T_c$}

Critical behavior of the asymmetry in $SU(2)$ gluodynamics
was studied in \cite{Bornyakov:2018mmf},
where the distribution of the configurations in the asymmetry 
was considered and the correlation between the asymmetry $\aasymt$ and 
the Polyakov loop ${\cal P}$ was found. 
Then the regression analysis based on the conditional cumulative distribution function $F({\cal A}|{\cal P})$ was employed to determine the dependence of 
the conditional expectation of the asymmetry
\beq
\langle {\cal A}\rangle_{\cal P}\equiv E({\cal A}|{\cal P})=
\int {dF({\cal A}|{\cal P})\over d{\cal A}} \; {\cal A} d{\cal A} 
\eeq
on the Polyakov loop, lattice volume, and the temperature.
It was found that, in the leading order in $\tau$, 
the volume and temperature dependence of the 
asymmetry is accounted for by its dependence on the Polyakov loop.

\begin{figure*}[tbh]
\vspace*{1mm}
\hspace*{-5mm}\includegraphics[width=9cm]{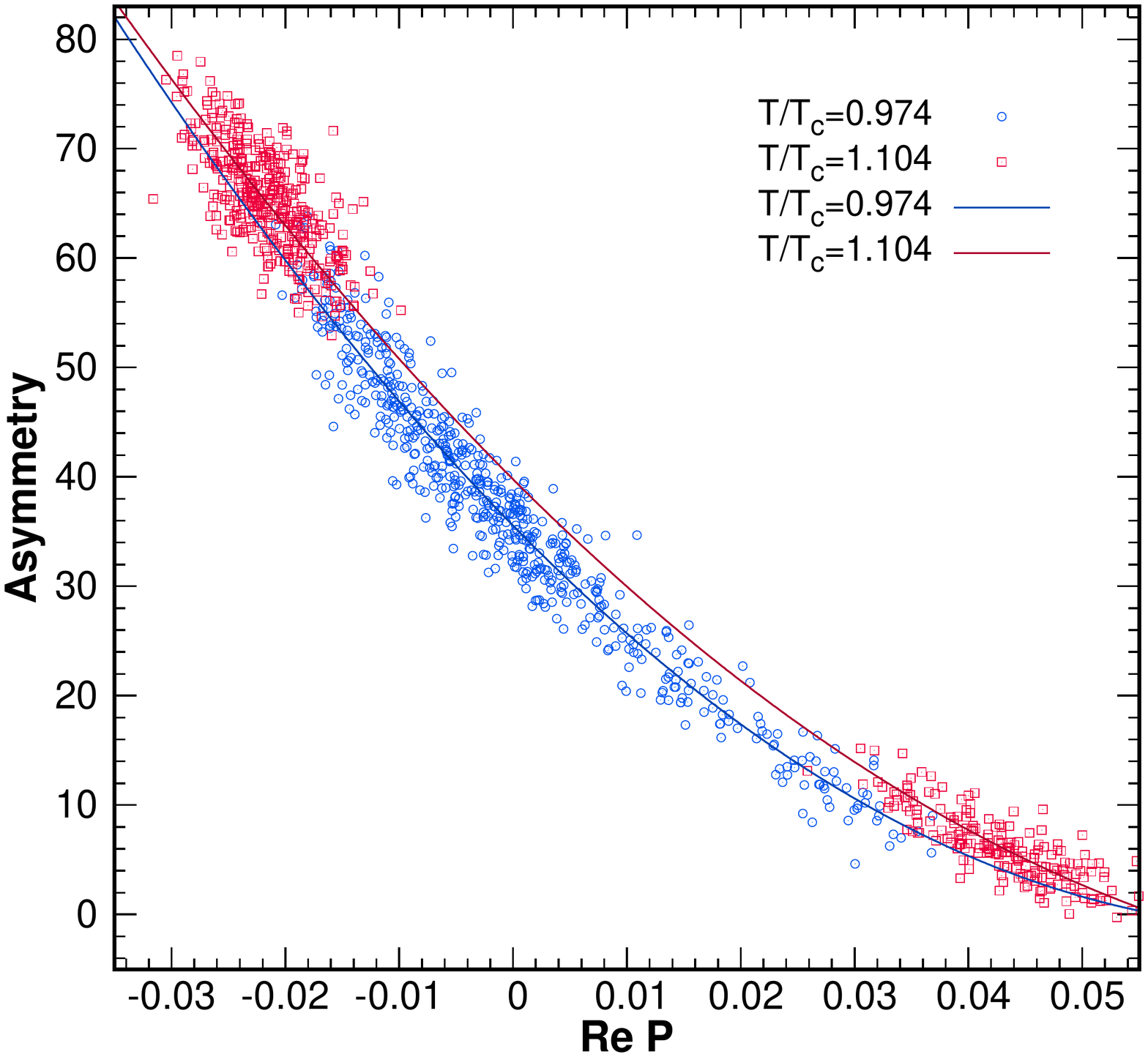}\hspace*{-8mm}
\includegraphics[width=9cm]{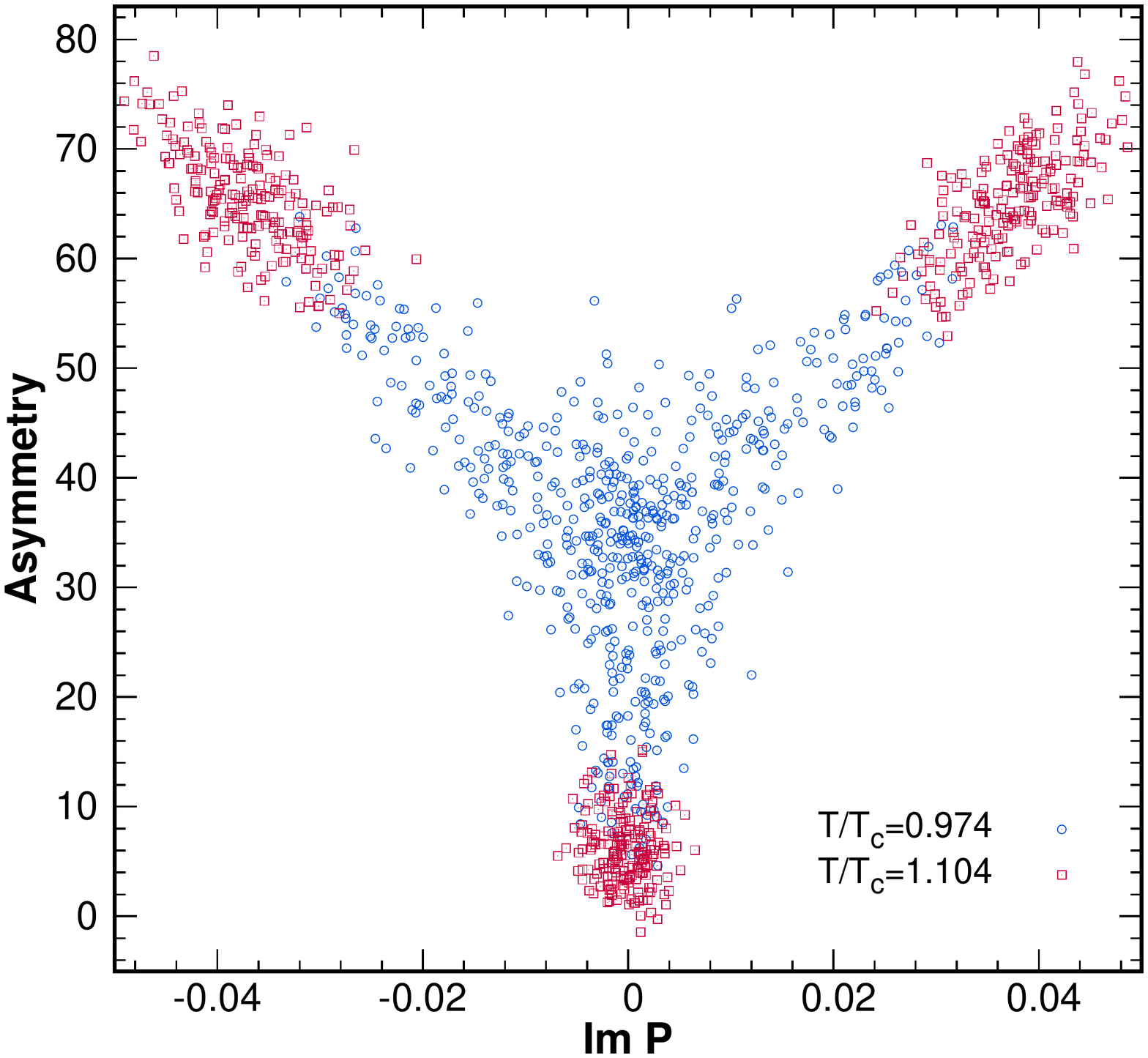}
\caption{Correlation between the asymmetry and 
the real part of the Polyakov loop (left);
scatter plot ``imaginary part of the Polyakov loop---asymmetry''
is consistent with the absence of correlation between them (right).}
\label{fig:asym_SU3}
\end{figure*}

In the $SU(3)$ case, the correlation between 
the asymmetry and the real part of the Polyakov loop 
is clearly seen on the scatter plot in the left panel 
of Fig.\ref{fig:asym_SU3}.
In view of this observation, we employ 
regression analysis to estimate a relationship 
between $\cal A$ and $\cal P$ using
the linear regression model based on the fit function
\beq
\label{eq:asym_regressed_fit}
E({\cal A}|\operatorname{Re}{\cal P}) 
\simeq \aasymt_0 + \aasymt_1 \operatorname{Re}{\cal P} + \aasymt_2 \big(\operatorname{Re}{\cal P}\big)^2 \,,
\eeq
where $\cal A$ is a predicted variable (regressand)
and $\operatorname{Re}{\cal P}$ is an explanatory variable (regressor).
The parameters $\aasymt_0$, $\aasymt_1$, and $\aasymt_2$ 
extracted from our data are presented in \Tab{tab:asym_regressed_fit}.
The residuals 
\beq\label{eq:}
e_n = {\cal A}_n - \aasymt_0 - \aasymt_1 \operatorname{Re}{\cal P}_n - \aasymt_2 (\operatorname{Re}{\cal P}_n)^2 \,,
\eeq
where subscript $n$ numbers gauge-field configurations,
show correlation with neither $\operatorname{Re}{\cal P}$ nor 
$\operatorname{Im}{\cal P}$. 
Hence our data give no evidence for a correlation between 
$\aasymt$ and $\operatorname{Im}{\cal P}$ 
or for a correction to the relation (\ref{eq:asym_regressed_fit}) 
between $\aasymt$ and $\operatorname{Re}{\cal P}$.

We have more to say on the temperature dependence 
of the asymmetry. In the infinite-volume limit, 
the width of the distribution of field configurations 
in the Polyakov loop tends to zero and, consequently, 
${\cal P}_n=\langle {\cal P}\rangle $\;. 
Thus the expectation value 
$ E\big({\cal A}\;\big|\;{\cal P}\!=\! {\cal P}(\tau)\big)$ 
determines the asymmetry  
in the infinite-volume limit:
\bea
\hspace*{-8mm} \langle \aasymt \rangle & = &  \aasymt_0(\tau)  \hspace*{35mm} \mbox{if} \quad \tau<0\;,\\ \nonumber
\hspace*{-8mm} \langle \aasymt \rangle & = & \aasymt_0(\tau) + \aasymt_1(\tau) \operatorname{Re}{\cal P}(\tau) \;+  \\ \nonumber
&& \;+\; \aasymt_2(\tau) (\operatorname{Re}{\cal P}(\tau))^2 \hspace*{13mm} \mbox{if} \quad \tau > 0\;. \nonumber
\eea
The coefficients $\aasymt_i$ evaluated on the lattices 
under consideration show rather smooth dependence 
on $\tau $ in a neighborhood of the point $\tau=0$ 
associated with the deconfinement transition: say, 
$\aasymt_0$ changes by some 5\% as $\tau$ changes 
from $-0.1$ to $0.0$\;. In the $SU(2)$ case, they 
not only posses this property but also depend 
very weakly on the lattice volume~\cite{Bornyakov:2018mmf}.
For this reason, it is natural to assume that 
the lattice size $\sim 2$~fm used in our study 
is sufficiently large for their evaluation.

As in the $SU(2)$ case, now we employ our knowledge of the 
critical behavior of the Polyakov loop for the 
investigation of the critical behavior of the asymmetry.
At $\tau>0$ spontaneous breaking of the 
center symmetry occurs and we choose a certain 
Polyakov-loop sector. In the infinite-volume limit,
${\cal P}$ is some function of $\tau$ such that
\beq
\lim_{\tau\to 0_+}|{\cal P}(\tau)| = {\cal P}_c > 0\;.
\eeq
The discontinuity ${\cal P}_c > 0$ 
implies that 
\beq
E\big({\cal A}\;\big|\;{\cal P}={\cal P}_c\big)\,-\, E\big({\cal A}\;\big|\;{\cal P}=0\big)\;=\;G_{\aasymt}^+ < 0\;,
\eeq
when we choose the Polyakov-loop sector with 
$\arg {\cal P}=0$ 
and  
\bea
&& E\big({\cal A}\;\big|\;{\cal P}=e^{2\imath \pi\over 3}{\cal P}_c\big)\, -\, E\big({\cal A}\;\big|\;{\cal P}=0\big)\;= \\ \nonumber  
=&& \hspace*{-2mm} E\big({\cal A}\;\big|\;{\cal P}=e^{-\,2\imath \pi\over 3}{\cal P}_c\big)\, -\, E\big({\cal A}\;\big|\;{\cal P}=0\big)\;=\; G_{\aasymt}^- > 0 \nonumber
\eea
otherwise. That is, discontinuity in the Polyakov loop at $T=T_c$ 
gives rise to the discontinuity of the asymmetry.

Our regression analysis indicates that the dependence of 
$ \aasymt$ on $\operatorname{Re}\!{\cal P}$ 
is much stronger than on 
$\operatorname{Im}\!{\cal P}$ and $\tau $;
that is, temperature dependence of $\cal A$ 
at $\tau > 0$ is accounted for mainly by  $\operatorname{Re}\!{\cal P}$. 
Scatter plot in the right panel of 
Fig.\ref{fig:asym_SU3} demonstrates that the values of $\aasymt$
plotted against $\operatorname{Im}\!{\cal P}$ look like 
the values of $\operatorname{Re}\!{\cal P}$ plotted against 
$\operatorname{Im}\!{\cal P}$.
Such pattern agrees well with the conclusion that
$\aasymt$ is independent of $\operatorname{Im}\!{\cal P}$.

\begin{table}[tbh]
\begin{center}
\vspace*{0.2cm}
\begin{tabular}{|c|c|c|c|} \hline
      &            &             &              \\[-2mm]
$\tau$ & ~~ $A_0$~~ & ~~ $A_1$~~  & ~~ $A_2$~~  \\[-2mm]
      &            &             &              \\
   \hline\hline
-0.096 & 33.98(14) & -962.8(19.4)   &  -\,603(2182)  \\
-0.026 & 35.54(14) & -1060.5(13.5)  &  7645(644)  \\
 0.025 & 37.26(24) & -1104.6(8.7)   &  7773(393)  \\
 0.104 & 39.78(35) & -1040.3(8.5)   &  5969(342)  \\
\hline\hline
\end{tabular}
\end{center}
\caption{Results of the fit (\ref{eq:asym_regressed_fit}).
}
\label{tab:asym_regressed_fit}
\end{table}

\section{Gluon propagators near criticality}

\begin{figure*}[hht]
\vspace*{1mm}
\hspace*{-5mm}\includegraphics[width=9cm]{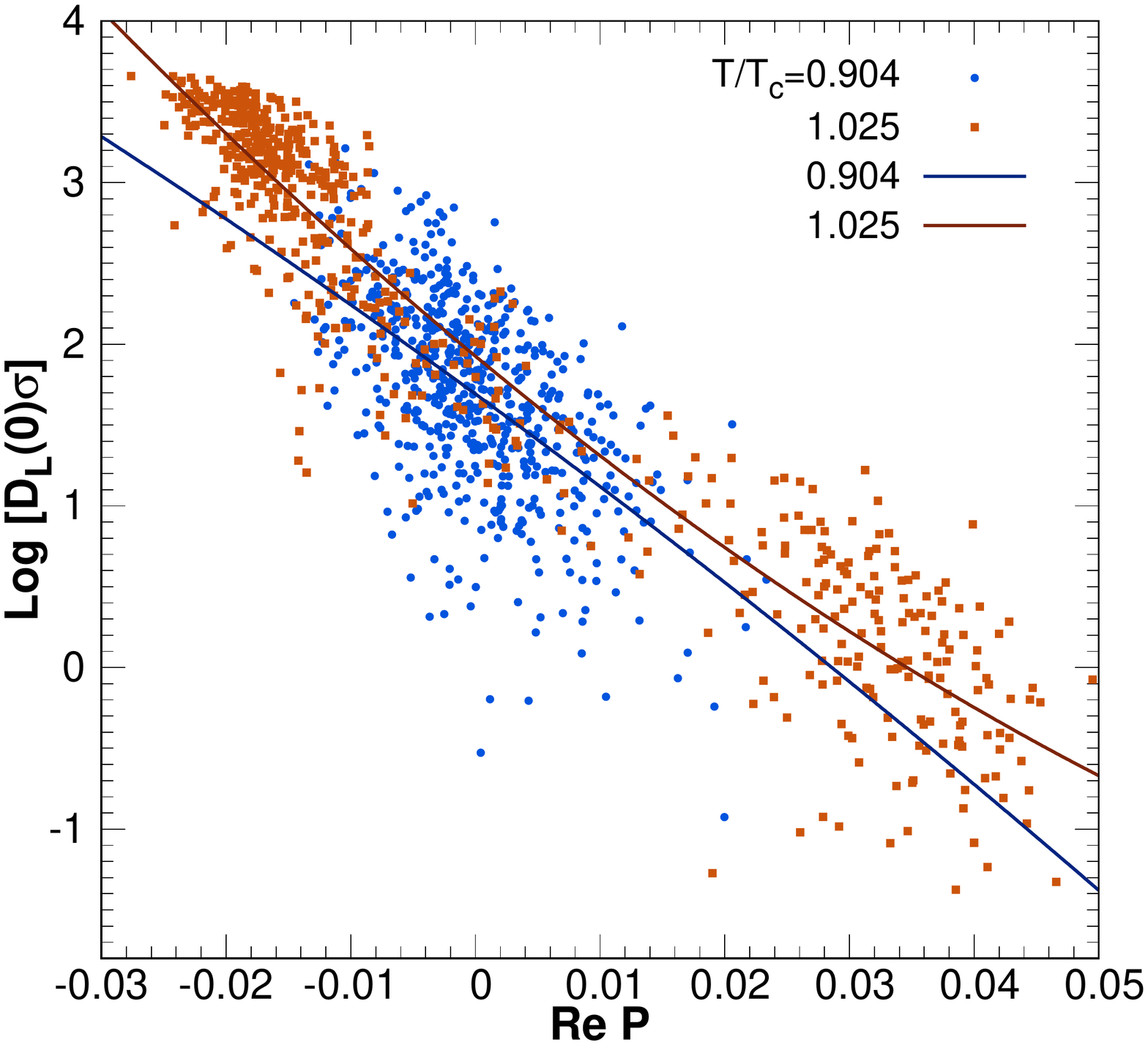}\hspace*{-8mm}
\includegraphics[width=9cm]{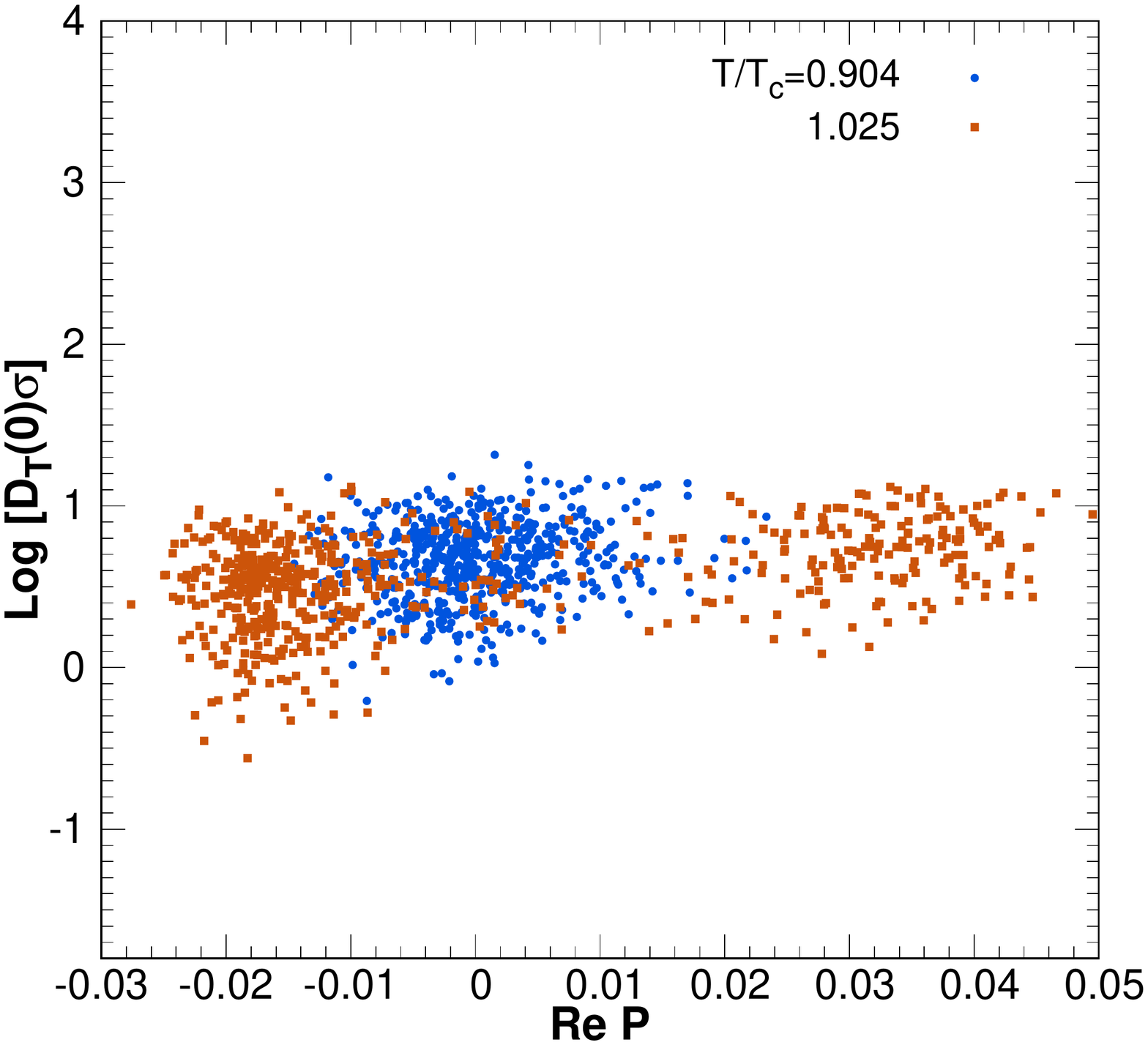}
\caption{Correlation between the longitudinal (left)
and transverse (right) gluon propagator at zero momentum 
and the real part of the Polyakov loop.}
\label{fig:glpr_ploo_SU3}
\end{figure*}

We begin with the observation that the zero-momentum
longitudinal propagator 
is strongly correlated with the real part of the Polyakov loop, 
see the scatter plot in Fig.\ref{fig:glpr_ploo_SU3}.
Some correlation between $D_T(0)$ and $\operatorname{Re} {\cal P}$
also takes place, whereas neither $D_L(0)$ nor $D_T(0)$ 
has a correlation with $\operatorname{Im} {\cal P}$.

We prove this relying on a procedure analogous to that
used in the case of asymmetry. Namely, we also begin with
the  conditional distribution of the propagator values
and find the average value of the propagator 
as a function of the Polyakov loop using a linear regression model.
An important difference from the case of asymmetry is that 
homoscedasticity of conditional distributions in $D_L(0)$
at various values of $\cal P$ is severely broken.
The heteroscedasticity is so great that it can hardly be evaluated 
on the basis of our limited data set. This stems mainly 
from non-Gaussian character of the 
distribution of configurations in $D_L(0)$,
which also holds for the conditional distributions 
at fixed ${\cal P}$. To obviate this problem, we consider
the quantity
\beq
{\cal D} = \ln \big(D_L(0)\sigma\big)
\eeq
such that the conditional distributions of configurations 
in it are normal (at least approximately) and 
the heteroscedasticity can be evaluated.
Having such evaluation, we find the conditional average
\beq
\langle {\cal D}\rangle_{\cal P}\equiv E({\cal D}|{\cal P})=
\int {dF({\cal D}|{\cal P})\over d{{\cal D}}} \; {\cal D} d{\cal D}\;,
\eeq
where $F({\cal D} |{\cal P})$ is the 
cumulative distribution function of ${\cal D}$   
at a given value of the Polyakov loop ${\cal P}$. 
For this purpose, we employ the linear 
regression model based on the fit function
\beq\label{eq:DL_regressed_fit}
{\cal D} \simeq {\cal D}_0(\tau) 
+ {\cal D}_1(\tau) \operatorname{Re}{\cal P}(\tau)
+ {\cal D}_2(\tau) \big(\operatorname{Re}{\cal P}(\tau)\big)^2.
\eeq
The results of our analysis are presented in~\Tab{tab:DL_regressed_fit}.
Over the range $-0.1 < \tau < 0.1$ variation of 
${\cal D}(\tau)$ caused by the change of the coefficients 
${\cal D}_0$ and ${\cal D}_1$ is much smaller than the variation
caused by the change of ${\cal P}$ according to formula 
(\ref{eq:DL_regressed_fit}), whereas the coefficient 
${\cal D}_2$ is poorly determined on our statistics.
Residuals show correlation with neither $\operatorname{Re}{\cal P}$
nor $\operatorname{Im}{\cal P}$ indicating that
$D_L(0)$ does not depend on $\operatorname{Im}{\cal P}$
and, within precision available on our data, 
its dependence on $\operatorname{Re}{\cal P}$
is accounted for by formula (\ref{eq:DL_regressed_fit})
with the coefficients from \Tab{tab:DL_regressed_fit}.

In \cite{Maas:2011ez} it was concluded 
on the basis of simulations 
on the $34^3 \times 4$ lattice 
that a pronounced jump 
of the longitudinal propagator 
is formed at the transition,
however, it is not clear whether 
this discontinuity survives 
the infinite-volume limit.
We argue that such jump can be 
caused by the difference
of the average values of the longitudinal propagator
in different Polyakov-loop sectors.
In a sufficiently small volume, this difference 
may be substantial even at $T<T_c$ (see next subsection for more detail);
therewith, in a finite volume
the center symmetry is not broken 
and no discontinuities in the propagator should emerge.
Nevertheless, one can see a discontinuity
in the temperature dependence of the propagator
even on a small-size lattice
at any given temperature $T_{fake}$
provided that one takes into account all three
Polyakov-loop sectors at $T<T_{fake}$ and 
only one Polyakov-loop sector at $T>T_{fake}$.
Obviously, such a jump is unrelated to the phase transition.
The proper  jump of the longitudinal propagator should appear 
only in the infinite-volume limit and we argue that such a jump
does emerge.

\begin{table*}[htb]
\begin{center}
\vspace*{0.2cm}
\begin{tabular}{|c|c|c|c|} \hline
      &            &             &              \\[-2mm]
$\tau$ & ~~ ${\cal D}_0\pm \delta_{STAT} \pm \delta_{SYST}$~~ & ~~ ${\cal D}_1\pm \delta_{STAT} \pm \delta_{SYST}$~~  & ~~ ${\cal D}_2\pm \delta_{STAT} \pm \delta_{SYST}$~~  \\[-2mm]
      &            &             &              \\
   \hline\hline
-0.096 & $1.692\pm 0.024 \pm 0.002$ & $-56.2 \pm 3.5 \pm 0.1$ &  $-94 \pm 374 \pm 10    $ \\
-0.026 & $1.753\pm 0.026 \pm 0.001$ & $-66.7 \pm 1.9 \pm 0.5$ &  $293 \pm 108 \pm 11$ \\
 0.025 & $1.926\pm 0.038 \pm 0.026$ & $-64.4 \pm 1.2 \pm 2.4$ &  $ 262 \pm 87 \pm 40$   \\
 0.104 & $2.240\pm 0.044 \pm 0.178$ & $-51.4 \pm 1.0 \pm 4.0$ &  $-157 \pm 42 \pm 95$  \\
\hline\hline
\end{tabular}
\end{center}
\caption{Results of the fit (\ref{eq:DL_regressed_fit}).
For each coefficient we present both the statistical error $\delta_{STAT}$  
and the systematic error $\delta_{SYST}$ associated with different methods of evaluation of the heteroscedasticity.
}
\label{tab:DL_regressed_fit}
\end{table*}

Our reasoning is based on
\begin{itemize}
 \item smooth dependence of the quantity
$\langle {\cal D}\rangle_{\cal P}$ (and, therefore, $D_L(0)$)
on the Polyakov loop
\item the fact that, in the infinite-volume
limit, the distribution in ${\cal P}$ becomes infinitely narrow
(the Polyakov-loop susceptibility tends to infinity as $V\to\infty$);
\item the assumption that the coefficients ${\cal D}_i$ in formula 
(\ref{eq:DL_regressed_fit}) depend on the lattice size only weakly 
(this does occur in the $SU(2)$ gluodynamics, however,
should be verified in the $SU(3)$ case).
\end{itemize}

Thus regression analysis gives some evidence that the dependence 
of ${\cal D}$ (and, therefore, of $D_L(0)$) on ${\cal P}$ 
near the criticality is rather smooth and it is 
reasonable to draw some consequences of this.
Having regard to the fact that, in the infinite-volume limit,
the Polyakov loop is a discontinuous function of the temperature,
we conclude that the zero-momentum longitudinal gluon 
propagator is also discontinuous. 

It should be noted that the quantity 
$\ds  {\exp \big( \langle {\cal D} \rangle \big) \over \sigma} $
gives a biased estimate of $\langle D_L(0)\rangle$.
However, here we focus only on qualitative reasoning and
it is sufficient for our purposes that this bias 
can in principle be evaluated and smooth dependence of 
$\langle{\cal D}\rangle_{\cal P}$ on ${\cal P}$ implies 
smoothness of $\langle D_L(0)\rangle_{\cal P}$.

Here one should make a comment 
on the temperature dependence 
of the zero-momentum longitudinal propagator shown 
in the left panel of Fig.2 in Ref.\cite{Aouane:2011fv}, 
where it is seen that $D_L(0)$ increases with temperature 
at $0< T < T_x$ and approaches its peak at some $T_x$ 
such that $T_x<T_c$.
When $|\tau| \sim 1$, this growth is unrelated to the correlation 
with the Polyakov loop because the values of ${\cal P}$ are distributed
close to zero and we ignore a reason of such growth.
However, at subcritical temperatures ($ \tau < 0, \ |\tau| <\!\!\!< 1$)
the width of the $|{\cal P}|$ distribution begins to rise. 
It should be emphasized that, at a nonvanishing width of 
the $|{\cal P}|$ distribution, $D_L(0)$ depends substantially on 
which Polyakov-loop sectors are taken into consideration.

Since only sector ($I$) was taken into account in~Ref.\cite{Aouane:2011fv},
the correlation shown in Fig.\ref{fig:glpr_ploo_SU3} indicates that
a broadening of the $|{\cal P}|$ distribution results in a decrease of $D_L(0)$.
However, in another Polyakov-loop sector, $D_L(0)$ 
increases with broadening of the $|{\cal P}|$ distribution.
When all three sectors are taken into consideration,
$D_L(0)$ continues to increase up to $T=T_c$ and 
decreasing of $D_L(0)$ with temperature at $T>T_c$
occurs only due to the restriction to sector $(I)$.

In a finite volume both the Polyakov loop and the propagator
are smooth functions of $T$. However, when one takes an average
$\langle D_L(0) \rangle$ over configurations in all three sectors at $T<T_c$
and over configurations only in sector $(I)$ at $T>T_c$, the above reasoning 
implies that such an average jumps down at $T=T_c$ 
and this is a fake discontinuity associated with 
an abrupt artificial restriction to only one sector:
it is precisely what was shown in \cite{Maas:2011ez}.
Such restriction is justified only in the infinite-volume limit.

In this limit, $D_L(0)$ jumps precisely at $T=T_c$ exactly
opposite to the Polyakov loop: it jumps down provided that sector 
$I$ is chosen by the system at $T>T_c$ and jumps up 
when the system at $T>T_c$ chooses sector $II$ or $III$.

In any case, for a comprehensive investigation 
of the critical behavior of Green functions
all Polyakov-loop sectors should be taken into account
in some neighborhood of the critical temperature.

\subsection{Propagators in different Polyakov-loop sectors}

Our statistics is not sufficient for a detailed study
of the gluon propagators at a given value of the Polyakov loop
or, more precisely, at a given value of 
$\operatorname{Re} {\cal P}$ since they are independent of 
$\operatorname{Im} {\cal P}$. However, certain conclusions
on the behavior of the propagators below
and above $T_c$ can be drawn simply from considering 
them in different Polyakov-loop sectors. Since the propagators 
computed in sector ($II$) coincide with those in sector ($III$),
we compare the propagators evaluated in sector ($I$)
referred to as ``Re~${\cal P}>0$'' with those evaluated on configurations from sectors ($II$) and ($III$) referred to as 
``Re~${\cal P}<0$''.

We begin with a comparison of zero-momentum values of the 
propagators presented in \Tab{tab:Dzero_in_sectors_I-III}.
In addition to the rapid decrease with increasing Polyakov loop
shown in the left panel of Fig.\ref{fig:glpr_ploo_SU3}, 
it is clearly seen in \Tab{tab:Dzero_in_sectors_I-III} 
that the zero-momentum longitudinal propagator 
decreases with temperature at Re~${\cal P}>0$
and increases with temperature at Re~${\cal P}<0$.
For the lattice size under consideration ($\sim 2$~fm)
the ratio 
$$
j={D_L^{\operatorname{Re}{\cal P} <0}(0)\over D_L^{\operatorname{Re}{\cal P} >0}(0)
}
$$ 
runs up to 3 well below the critical temperature ($\tau=-0.026$).
Above the critical temperature, $j$ shows a rapid growth and reaches
30 at $\tau \approx 0.1$.

\begin{table}[htb]
\begin{center}
\vspace*{0.2cm}
\begin{tabular}{|c|c|c|c|c|} \hline
       &                          &                          &      
                         &                          \\[-2mm]
$\tau$ & ~~$D_L(0) $~~ & ~~$D_L(0)$~~ & ~~$D_T(0)$~~ & ~~$D_T(0)$ ~~ \\[2mm]
       &  $\operatorname{Re}{\cal P}>0$  & $\operatorname{Re}{\cal P}<0$  &  $\operatorname{Re}{\cal P}>0$ &  $\operatorname{Re}{\cal P}<0$   \\    \hline\hline
-0.096 & 20.88(65)  & 36.16(1.17)  & 9.36(14) & 8.79(12)  \\
-0.026 & 17.7(1.3)  & 54.97(1.93)  & 9.19(16) & 8.35(12)  \\
 0.025 & 8.82(1.22) & 99.35(2.11)  & 9.44(14) & 7.57(11)  \\
 0.104 & 3.95(16)   & 125.53(1.38) & 9.30(15) & 6.68(10)  \\
\hline\hline
\end{tabular}
\end{center}
\caption{Average values of the zero-momentum propagators 
in different Polyakov-loop sectors. No difference between sectors 
($II$) and ($III$) has been found, they are referred to 
as ``$\operatorname{Re}{\cal P}<0$''.}
\label{tab:Dzero_in_sectors_I-III}
\end{table}

\begin{figure*}[tbh]
\vspace*{1mm}
\hspace*{-5mm}\includegraphics[width=9cm]{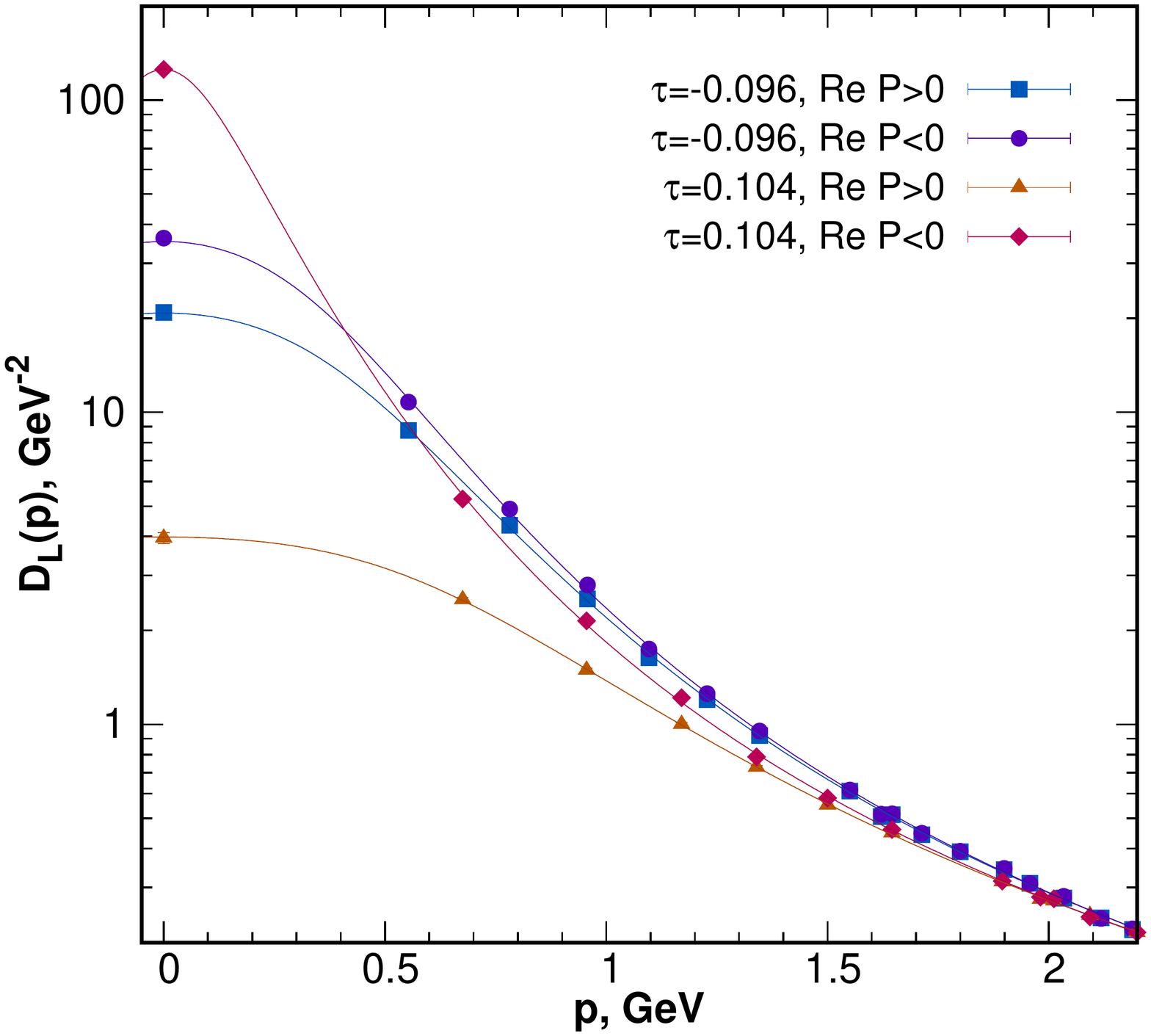}\hspace*{-8mm}
\includegraphics[width=9cm]{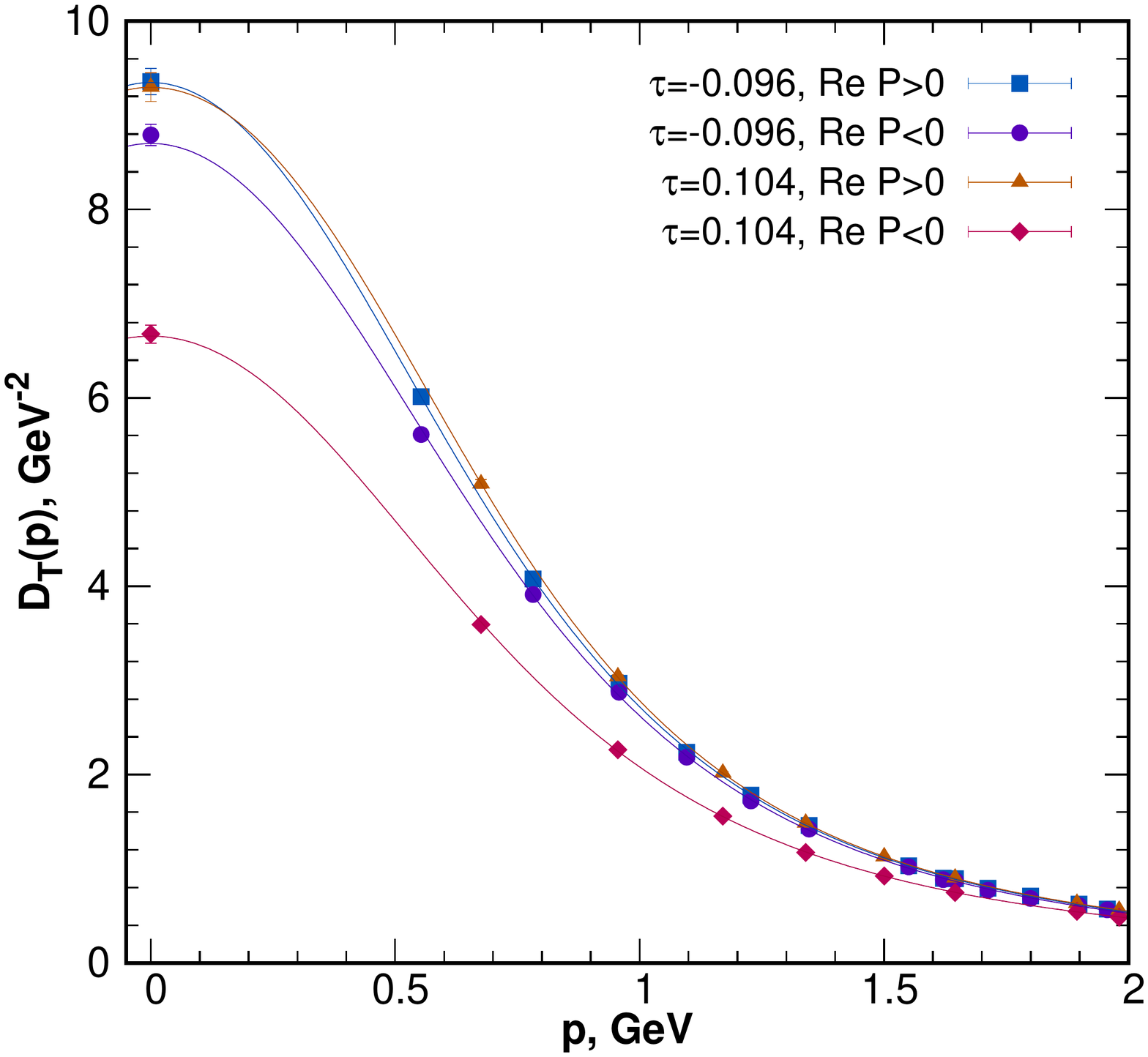}
\caption{Longitudinal (left)
and transverse (right) gluon propagator 
as functions of the momentum 
in different Polyakov-loop sectors below and above critical temperature.
Notice logarithmic scale on the ordinate axis on the left panel.}
\label{fig:glpr_vs_mom}
\end{figure*}

Both in the right panel of Fig.\ref{fig:glpr_ploo_SU3} 
and in \Tab{tab:Dzero_in_sectors_I-III}
it is demonstrated that, in contrast to the 
longitudinal propagator, the zero-momentum transverse 
propagator slightly increases 
with an increase of the Polyakov loop.
Within statistical errors, it does not 
change with temperature at  Re~${\cal P}>0$
and shows moderate decreasing at  Re~${\cal P}<0$.

The propagators at momenta $p<2.2$~GeV for different
Polyakov-loop sectors are shown in Fig.\ref{fig:glpr_vs_mom}.

The curves in Fig.\ref{fig:glpr_vs_mom} are the results 
of the fit based on the Gribov-Stingl formula
\beq
D(p)\simeq c {p^2+d^2\over (p^2+M^2)^2+b^4 } 
\eeq
over the momentum range $p<2.5$~GeV,
here we only mention that it works well.
The transverse propagator shown in the right panel
shows a noticeable difference
$$
\mathbf{d}_T(p)=D_T^{\operatorname{Re}{\cal P} <0}(p)-D_T^{\operatorname{Re}{\cal P} >0}(p)
$$
which rapidly decreases with increasing momentum and moderately 
increases with increasing temperature. This being so, the transverse
propagator is independent of temperature in the sector 
$\operatorname{Re}{\cal P} >0$ and is decreasing with 
temperature in the sector $\operatorname{Re}{\cal P} <0$.

As compared with the transverse propagator, the longitudinal propagator
features a huge difference 
$$
\mathbf{d}_L(p)=D_L^{\operatorname{Re}{\cal P} <0}(p)-D_L^{\operatorname{Re}{\cal P} >0}(p)
$$ 
in a deep infrared, which is, however, characterized by
a very sharp decrease with the momentum, especially at $\tau > 0$.
Though $\mathbf{d}_L(p)$ increases with increasing temperature, the
behavior of the longitudinal propagator by itself is more complicated.
In the sector $\operatorname{Re}{\cal P} >0$, the longitudinal propagator
substantially decreases with increasing temperature 
over the range $-0.1<\tau< 0.1$,
whereas in the sector $\operatorname{Re}{\cal P} <0$ it decreases 
with increasing temperature only at $p\gtrsim 0.4$~GeV; at smaller momenta 
it rapidly increases over the range $-0.1<\tau< 0.1$.

\subsection{Screening masses}

We also evaluate screening for different sectors of the Polyakov loop.
The chromoelectric and chromomagnetic screening masses 
are obtained from the fit of the formula 
\beq\label{eq:small_mom_ff}
{1 \over D_{L,T}(p)} \simeq {1\over Z} \left(m_{E,M}^2 + p^2 + r p^4\right) 
\eeq 
to the data on inverse longitudinal and transverse 
gluon propagators at low momenta, respectively; 
for more detail about this definition of screening 
masses see Refs.\cite{Bornyakov:2019jfz,Bornyakov:2020kyz}.
We performed fit over the range $0\leq p < 1.3$~~GeV,
in this domain the fit formula (\ref{eq:small_mom_ff}) 
works well for both the longitudinal and the transverse propagators
giving a reliable fit quality for all Polyakov-loop sectors. 
The fit is stable with respect to variations of the upper 
bound of the fit range between 1.2 and 1.6~GeV and 
to an exclusion of zero momentum.
The results for screening masses are presented 
in \Tab{tab:scree_mass_in_sect_I-III}.

Chromomagnetic screening masses in different Polyakov-loop sectors 
differ slightly, if at all, and their temperature dependence is 
not clearly seen. For this reason we take a closer look at 
chromoelectric sector and consider $m_E$ and $D_L(0)$ 
relative to their chromomagnetic counterparts.

\begin{table}[tbh]
\begin{center}
\vspace*{0.2cm}
\begin{tabular}{|c|c|c|c|c|} \hline
       &                          &                          &      
                         &                          \\[-2mm]
$\tau$ & ~~$m_E^2$~~ & ~~$m_E^2$~~ & ~~$m_M^2$~~ & ~~$m_M^2$ ~~ \\[2mm]
       &  $\operatorname{Re}{\cal P}>0$  & $\operatorname{Re}{\cal P}<0$  &  $\operatorname{Re}{\cal P}>0$ &  $\operatorname{Re}{\cal P}<0$   \\    \hline\hline
-0.096 & 0.373(31) & 0.214(31)  & 0.638(34) & 0.642(39) \\
-0.026 & 0.445(71) & 0.136(11)  & 0.609(24) & 0.586(32) \\
 0.025 & 0.523(56) & 0.0498(38) & 0.672(37) & 0.565(18) \\
 0.104 & 0.95(20)  & 0.0272(11) & 0.664(43) & 0.611(8)  \\
\hline\hline
\end{tabular}
\end{center}
\caption{Values of the chromoelectric and chromomagnetic 
screening masses (in GeV$^2$) obtained 
by the fit formula (\ref{eq:small_mom_ff})
in different Polyakov-loop sectors. 
No difference between sectors 
($II$) and ($III$) has been found, they are referred to 
as ``$\operatorname{Re}{\cal P}<0$''.}
\label{tab:scree_mass_in_sect_I-III}
\end{table}

As $T/T_c$ increases from 0.9 to 1.1, the 
chromoelectric screening mass in the sector 
$\operatorname{Re}{\cal P}<0$ increases from
600~MeV to 1~GeV, whereas 
in the sector $\operatorname{Re}{\cal P}<0$
it decreases from 460~MeV to 160~MeV. 
The difference between the screening masses 
in different Polyakov-loop sectors at $T<T_c$
can be attributed to finite-volume effects
related to a finite width of the distribution
in $\operatorname{Re}{\cal P}$.
In the deconfinement phase 
screening of color charges is different 
in different sectors. At $T = 1.1T_c$, as an example,
screening radii differ substantially:
1.2~fm in the sector $\operatorname{Re}{\cal P}>0$
versus 0.2~fm in the sector $\operatorname{Re}{\cal P}<0$.

It was shown in \cite{Bornyakov:2021arl} that, 
when the screening mass is sufficiently large, 
the strength of chromoelectric or chromomagnetic interactions
between static color charges or currents 
should be characterized by the quantity
\beq\label{eq:VEM_definition}
V_E = m_E^3 D_L(0)\qquad \mbox{or} \qquad V_M = m_M^3 D_T(0)
\eeq
rather than by $D_L(0)$ or $D_T(0)$, respectively.
The screening masses in the case under consideration
are rather small, nevertheless, in~\Tab{tab:V_EM} we present the values 
of $V_{E,M}$ which have the meaning of the depth 
of the static-quark potential well 
in the limit of large screening masses. 
Though these quantities are normalization dependent, 
their ratios give information, in particular, 
on the strength of chromoelectric interactions 
relative to the strength of the chromomagnetic interactions in 
different Polyakov-loop sectors.

As the temperature increases from $0.9 T_c$ to $1.1 T_c$, 
the ratio $\ds \left. { V_E \over V_M}\right|_{\operatorname{Re}{\cal P}>0}$ decreases by some 30\%
and the ratio $\ds \left.{ V_E \over V_M}\right|_{\operatorname{Re}{\cal P}<0}$ decreases more than by a factor of four. 

\begin{table}[tbh]
\begin{center}
\vspace*{0.2cm}
\begin{tabular}{|c|c|c|c|c|} \hline
 & \multicolumn{2}{c}{} & \multicolumn{2}{|c|}{}    \\[-2mm]
$\tau$  & \multicolumn{2}{c}{$V_E$,~GeV} & \multicolumn{2}{|c|}{$V_M$,~GeV} \\[2mm]
\hline
      &  $\operatorname{Re}{\cal P}>0$  & $\operatorname{Re}{\cal P}<0$  &  $\operatorname{Re}{\cal P}>0$ &  $\operatorname{Re}{\cal P}<0$   \\    \hline\hline
-0.096 & 4.8 & 3.5  & 4.8 & 4.5 \\
-0.026 & 5.2 & 2.8  & 4.4 & 3.7 \\
 0.025 & 3.3 & 1.1  & 5.2 & 3.2 \\
 0.104 & 3.7 & 0.56 & 5.0 & 3.2  \\
\hline\hline
\end{tabular}
\end{center}
\caption{Strength of the chromoelectric ($V_E$)  and chromomagnetic 
($V_M$) interactions determined by formula (\ref{eq:VEM_definition}) in different Polyakov-loop sectors. Errors are not shown because the presented values can be used only for rough qualitative estimates.}
\label{tab:V_EM}
\end{table}

Thus in the deconfinement phase the relative strength
of chromoelectric interactions decreases slightly in the sector
$\operatorname{Re}{\cal P}>0$ and significantly in the sector $\operatorname{Re}{\cal P}<0$. This being so, the chromoelectric screening radius
decreases in the sector $\operatorname{Re}{\cal P}>0$ both in absolute value
and with respect to the chromomagnetic screening radius 
and dramatically increases in the sector $\operatorname{Re}{\cal P}<0$. 

Gluon matter in the deconfinement phase can be considered 
as chromomagnetic medium, that is, as a medium with weaken 
and well-screened chromoelectric interactions provided that the 
the sector $\operatorname{Re}{\cal P}>0$ is chosen.
Choosing one of the sectors with $\operatorname{Re}{\cal P}<0$,
we arrive at a medium with strong and short-range 
chromomagnetic interactions and weak and long-range 
chromoelectric interactions.

\subsection{Speculations on the deconfinement phase transition}

Thus the longitudinal propagator in the sector 
$\operatorname{Re}{\cal P} >0$ differs dramatically 
from the longitudinal propagator in the 
sector $\operatorname{Re}{\cal P} <0$.
In this connection, it is reasonable to recollect 
the confinement scenario proposed, in particular, in Refs.\cite{Fortunato:1999wr,Gattringer:2010ug} and investigated in \cite{Ivanytsky:2016bhb}.
In these works, the properties of the gluon medium 
responsible for confinement of heavy static quarks
were discussed. Such medium can be characterized in terms of
center clusters (the domains where the Polyakov loop
takes the values mainly from one sector).

In the deconfinement phase, there exists a percolating cluster
associated with some center element of the gauge group,
and the remaining space is either occupied  by finite-size clusters
associated with some center element or characterized 
by the values of the Polyakov loop that does not
clearly favor a definite center element.
As the temperature decreases, the part of space 
occupied by the percolating cluster decreases
until it disappears at the critical temperature.

Let us proceed to some qualitative speculations to outline 
directions of further investigations.
Our finding that the Polyakov-loop sectors 
differ in the infrared behavior of the longitudinal
gluon propagator gives some evidence that 
static color charges interact differently
in different clusters. Thus the Polyakov-loop sectors 
are not equivalent for gauge-dependent quantities
even in a pure gauge theory.
In the Landau gauge, we obtain that 
the ``trivial'' sector $\operatorname{Re}{\cal P} >0$ 
is preferred in the sense that it features the 
most screened chromoelectric interaction
between color static charges. This is the 
most natural choice of the Polyakov-loop value 
in the deconfinement phase. In the Landau gauge,
the finite-size clusters
associated with the other center elements
$\operatorname{Re}{\cal P} <0$ 
can be considered as ``bubbles of glue''
in the deconfinement phase:
in the Landau gauge, the longitudinal gluon propagator
provides long-range chromoelectric interaction of 
static color charges within such clusters.
Their volume increases with decreasing temperature, 
whereas the range of color-charge interaction 
within each such cluster decreases.
In a percolating cluster, the opposite happens. 
Its volume decreases with decreasing temperature,
whereas the range of chromoelectric forces increases.
At the critical temperature 
all clusters become identical (in the infinite-volume limit).

Such a scenario should be checked in further 
lattice simulations; in particular, the color-singlet
and color-octet potentials in each Polyakov-loop 
sector should be studied. Another problem
is to study the high-temperature behavior of
the differences $\mathbf{d}_L(p)$ and $\mathbf{d}_T(p)$ 
which rapidly decrease with increasing momentum. 
 It is interesting to find out whether 
there really is a momentum $p_J$, common to 
all temperatures, such that both
$\mathbf{d}_L(p)$ and $\mathbf{d}_T(p)$ 
become negligible at $p> p_J$.
Such a momentum would indicate
the boundary between nonperturbative infrared 
and perturbative ultraviolet domains
in gluodynamics.

\section{Conclusions}

We have studied the asymmetry $\aasymt$ and the 
longitudinal gluon propagator 
in the Landau-gauge $SU(3)$ gluodynamics on lattices $24^3\times 8$
over the range of temperatures $0.9 T_c < T < 1.1 T_c$. 
Our findings can be summarized as follows:

\begin{itemize}
\item Both the asymmetry $\aasymt$ and 
the zero-momentum longitudinal propagator $D_L(0)$ have a
significant correlation with the real part of the Polyakov loop ${\cal P}$.
The correlation between $D_T(0)$ and $\operatorname{Re}{\cal P}$
is non-negligible. Neither $\aasymt$ nor $D_L(0)$ nor $D_T(0)$
has a correlation with $\operatorname{Im}{\cal P}$.
\item We suggest a method to substantially reduce finite-volume effects. 
In the deconfinement phase, 
the conditional averages $\langle {\cal A} \rangle_{{\cal P}=z}$ 
or $\langle D_L(0) \rangle_{{\cal P}=z}$ give  
a close approximation to the infinite-volume limit
of $\aasymt$ or $D_L(0)$ at the temperature $\tau$
determined from the equation ${\cal P}_\infty(\tau)=z$ 
provided that $z$ is an allowed infinite-volume value 
of the Polyakov loop in a chosen sector and ${\cal P}_\infty$
is the infinite-volume expectation value of ${\cal P}$.
\item We determined critical behavior of  $\aasymt$ and $D_L(0)$
in the infinite-volume limit.
Regression analysis reveals that the conditional 
averages $\langle {\cal A} \rangle_{\cal P}$ 
and $\langle D_L(0) \rangle_{\cal P}$ 
are smooth functions of the Polyakov loop. 
Discontinuity in the Polyakov loop at $T=T_c$ 
in the infinite-volume limit
implies discontinuity of the asymmetry and 
the longitudinal gluon propagator.
The discontinuities of $\aasymt$ and $D_L(0)$
at $T=T_c$ are readily determined from the dependencies 
of $\langle {\cal A} \rangle_{\cal P}$ 
and $\langle D_L(0) \rangle_{\cal P}$ on $\operatorname{Re}\!{\cal P}$.
\item The infrared behavior of the longitudinal propagator
depends significantly on the Polyakov-loop sector; 
a moderate dependence of the transverse propagator 
in the infrared on the Polyakov-loop sector is also observed. 
\item In the deconfinement phase, distinctions between 
gauge-dependent quantities in different Polyakov-loop 
sectors are significant.
We have considered as an example
chromoelectric interactions 
relative to chromomagnetic interactions,
whose dependence on the temperature and the Polyakov-loop
sector is not very significant.
They  are weakly suppressed and short-range in  
the sector $\operatorname{Re}\!{\cal P}>0$
and moderately suppressed and long-range in  
each sector with $\operatorname{Re}\!{\cal P}<0$.
\end{itemize}

\vspace*{2mm}

\acknowledgments{Computer simulations were performed on the IHEP (Protvino),
Central Linux Cluster and ITEP (Moscow) Linux Cluster.
This work was supported by the Russian Foundation for Basic Research, 
grant~no.20-02-00737~A.}

\bibliographystyle{apsrev}
\bibliography{ref_crit_behavior}

\end{document}